\documentclass[
prl,
superscriptaddress, twocolumn,
amsmath,amssymb,
aps,
]{revtex4-2}

\UseRawInputEncoding 

\usepackage{graphicx}
\usepackage{color}
\usepackage{subfigure}
\usepackage{float}
\usepackage{amsmath}
\usepackage{ulem}
\usepackage{dcolumn}
\usepackage{bm}
\usepackage{hyperref}
\usepackage{CJKutf8}
\usepackage[utf8]{inputenc}

\newcommand{\yi}[1]{{\color{black}#1}}

\newcommand{\ket}[1]{|#1\rangle}
\newcommand{\bra}[1]{\langle#1|}
\newcommand{\vo}{\mathbf{0}}
\newcommand{\vx}{\mathbf{x}}
\newcommand{\vk}{\mathbf{k}}
\newcommand{\vq}{\mathbf{q}}
\begin{document}

\newcommand{\thistitle}{
Simulating Spin Dynamics of Supersolid States in a Quantum Ising Magnet
}

\title{\thistitle}
\author{Yi Xu (\begin{CJK*}{UTF8}{gbsn}徐熠\end{CJK*})}
 \email{slowlight@rice.edu}
\affiliation{Department of Physics and Astronomy, Rice University, Houston, Texas 77005, USA}

\author{Juraj Hasik}
 \email{j.hasik@uva.nl}
 \affiliation{Institute for Theoretical Physics and Delta Institute for Theoretical Physics, University of Amsterdam, Science Park 904, 1098 XH Amsterdam, The Netherlands}
 \affiliation{Department of Physics, University of Zurich, Winterthurerstrasse 190, 8057 Zurich, Switzerland}

\author{Boris Ponsioen}
 \email{b.g.t.ponsioen@uva.nl}
\affiliation{Institute for Theoretical Physics and Delta Institute for Theoretical Physics, University of Amsterdam, Science Park 904, 1098 XH Amsterdam, The Netherlands}

\author{Andriy H. Nevidomskyy}
 \email{nevidomskyy@rice.edu}
\affiliation{Department of Physics and Astronomy, Rice University, Houston, Texas 77005, USA}
\affiliation{Rice Center for Quantum Materials, Rice University, Houston, Texas 77005, USA}

\date{\today}

\begin{abstract}
Motivated by a recent experimental study on the quantum Ising magnet $\text{K}_2\text{Co}(\text{SeO}_3)_2$ that presented spectroscopic evidence of zero-field supersolidity~\href{https://arxiv.org/pdf/2402.15869.pdf}{(Chen \textit{et al}., arXiv:2402.15869)}, we simulate the excitation spectrum of the corresponding microscopic $XXZ$ model for the compound, using the recently developed excitation ansatz for infinite projected entangled-pair states. We map out the ground state phase diagram and compute the dynamical spin structure factors across a range of magnetic field strengths, focusing especially on the two supersolid phases found near zero and saturation fields. Our simulated excitation spectra for the zero-field supersolid ``Y" phase are in excellent agreement with the experimental data, recovering the low-energy branches and integer quantized excited energy levels $\omega_n=nJ_{zz}$. 
Furthermore, we demonstrate the nonlocal multi-spin-flip features for modes at $\omega_2$, indicative of their multi-magnon nature. Additionally, we identify characteristics of the high-field supersolid ``$\Psi$" phase in the simulated spectra, which should be compared with future experimental results.
\end{abstract}

\maketitle

\noindent
\textit{Introduction.}---
Supersolid (SS) is a unique quantum phase of matter, originally conjectured to exist in the phase diagram of $^4\text{He}$~\cite{chester70}, where both the global U(1) symmetry of the order parameter and the translation symmetry are spontaneously broken, and thus to possess both superfluidity and solidity.
While evidence in favor of a supersolid phase in helium remains controversial~\cite{tiwari2004supersolid, Day_2007, Kim_2012, Boninsegni_2012}, it has been proposed that supersolids and their lattice analogues can be realized in various other systems, ranging from ultracold atom platforms using atomic Bose-Einstein condensates~\cite{L_onard_2017, Li_2017, Guo_2019, Tanzi_2019, Bland_2022}, to triangular lattice hard-core boson systems~\cite{melko05, wessel05, boninsegni05, heidarian05, wang09}. 
The latter have a natural correspondence to spin-$1/2$ Heisenberg-type models on the triangular lattice via the mapping of the boson creation and annihilation operators onto the spin ladder operators via the Matsubara--Matsuda transformation~\cite{matsubara-matsuda}.
Recent experimental advances in  the synthesis and characterization of layered triangular lattice magnets based on cobalt~\cite{co_exp_PhysRevB.102.224430, exp_co_Kamiya_2018, chen2024phase, zhu2024continuum, Xiang2024} and ytterbium~\cite{qsl-exp-trgl-lat-NaYbO2,qsl-exp-trgl-lat-NaYbSe2, Xie2023} provide additional motivation for studying these  models theoretically.

The ground state phase diagrams of triangular lattice spin-$1/2$ models have been studied extensively using a variety of numerical techniques, from exact diagonalizations on small clusters to variational methods simulating lattices with hundreds of sites, in particular the density matrix renormalization group (DMRG) and variational Monte Carlo~\cite{ed_j1j2, yama14-xxz-cmf, zhang06-dmrg_trgl_xxz,li15-j1j2_trgl-cc,zhu15-j1j2_trgl-dmrg,hu15-j1j2_trgl-dmrg,saada16-j1j2_trgl-idmrg,mish13-j1j2_trgl-vmc,Kaneko14-j1j2_trgl-vmc,iqbal16-j1j2_trgl-vmc,bauer-j1j2_trgl-sbmf,zhu18-sl_trgl-dmrg,gong19-gs_dmrg_j1j2j3_trgl, ulaga24-ft-xxz-trgl}.
These variational methods have been further extended to simulate low-lying excitation spectra, with applications targeting the $J_1$-$J_2$ spin-1/2 model on the triangular lattice~\cite{ferrari19-exci_vmc_trgl_j1j2, drescher23-exci_dmrg_j1j2, sherman23-exci_dmrg_j1j2}. In contrast to finite-size approaches, infinite projected entangled-pair states (iPEPS) have been introduced to study two-dimensional systems directly in the thermodynamic limit~\cite{peps2006, jordan08, corboz14}, offering a systematically improvable ansatz which avoids finite-size effects.
Recently, the iPEPS ansatz was also extended to simulate the excited states of frustrated spin-1/2 systems and applied to the square~\cite{laurens15-exci,laurens19-exci,Ponsioen20-exci_peps,adpeps,tu24-exci_j1j2}, honeycomb~\cite{chen24-exci-kitaev,wang2024-exci-kitaev}, and triangular lattice models~\cite{chi22-exci_xxz_trgl}.
Despite these advances, theoretical understanding of the supersolid states on the triangular lattice in terms of the spin dynamics, which is beyond linear spin wave description, is still lacking.

In this Letter, motivated by the very recent evidence of the supersolid state from the inelastic neutron scattering data on the triangular lattice magnet $\text{K}_2\text{Co}(\text{SeO}_3)_2$~\cite{chen2024phase, zhu2024continuum}, we perform numerical iPEPS simulations of the spin dynamics in the XXZ spin-1/2 model with strong easy-axis exchange anisotropy.
We find that our zero-field spectra 
for the low-field SS ``Y" phase (SSY) agree well with the experimental neutron scattering data, corroborating its interpretations as a supersolid. 
The associated superfluid order parameter remains finite at small fields, while for zero field our results indicate possible restoration of the U(1) symmetry, also reported in the recent exact diagonalization study~\cite{ulaga2024easyaxisheisenbergmodeltriangular}.
The experimental magnetization data in strong magnetic fields ($B_z\geq20$~T) point toward the existence of another, high-field supersolid phase characterized by the ``$\Psi$"-shape of the spins on each triangle. Given that inelastic neutron scattering (INS) data at these very high magnetic fields are currently unavailable, we analyze this SS $\Psi$ phase (SSP) and compute its dynamical spin structure factor, making predictions that should be compared with future INS data for $\text{K}_2\text{Co}(\text{SeO}_3)_2$ or other triangular lattice compounds.

\begin{figure}[thb]
    \flushleft
    \includegraphics[width=0.45\textwidth]{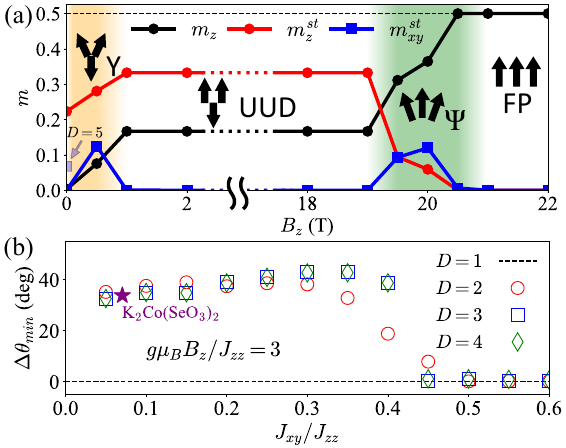}
    \caption{(a) Ground state phase diagram for K$_2$Co(SeO$_3$)$_2$, obtained from iPEPS calculations with the finite bond dimension scaling up to $D=5$.
    Two identified supersolid phases, the Y phase  and $\Psi$ phase, are highlighted in orange and green, respectively. Both have non-zero staggered transverse magnetization ($m^{st}_{xy}=|\frac{1}{3}\sum_{n\in u.c.}e^{2in\pi/3}[\langle S_n^x\rangle \hat{x}+\langle S_n^{y}\rangle\hat{y}]|$) and non-uniform out-of-plane staggered magnetization ($m_z^{st}=|\frac{1}{3}\sum_{n\in u.c.}e^{2in\pi/3}\langle S_n^z\rangle|$) in the field direction ($\hat{z}$)~\cite{murakami-ss-order-param}, interpreted as spontaneous breaking of U(1) global symmetry and \yi{a $C_3$ lattice symmetry}, respectively. The black line corresponds to the uniform magnetization, $m_z=|\frac{1}{3}\sum_{n\in u.c.}\langle S_n^z\rangle|$. (b) The minimal mutual angle $\Delta\theta_{min}$ for different $J_{xy}/J_{zz}$ at fixed $g\mu_B B_z/J_{zz}=3$. }
    \label{fig:phase_diagram}
\end{figure}

\vspace{1mm}
\noindent
\textit{Model and methods.---} We study the $S=1/2$ XXZ model with easy-axis exchange anisotropy under uniform magnetic field on the triangular lattice, whose microscopic Hamiltonian is given by (setting $\hbar=1$ throughout)
\begin{equation}
    \hat{H}= \sum_{\langle ij\rangle} \big [J_{zz} S^z_i S^z_j + J_{xy}\big(S^x_i S^x_j + S^y_i S^y_j \big)\big] - g\mu_B B_z \sum_{i} S_i^z,
    \label{eq:model}
\end{equation}
where $J_{zz}$ and $J_{xy}$ are the exchange coupling strengths between the corresponding spin components of the nearest neighbor $\langle ij\rangle$ spins on the triangular lattice, and $g$ is the Land\'e $g$-factor with $\mu_B$ being the Bohr magneton. 
In our simulations, we fix the exchange coupling and XXZ anisotropy by adopting the experimentally determined values in Ref.~\cite{chen2024phase}: $J_{zz}=2.98$~meV, $J_{xy}=0.21$~meV, and $g=7.8$.  
To describe the three-sublattice (``$abc$'') order on the triangular lattice, we parametrize ground states by iPEPS $|\Psi(A)\rangle$ with a unit cell consisting of three site tensors, repeated on infinite lattice
\begin{equation}
|\Psi(A=[a,b,c])\rangle=\sum_{\{s\}}\vcenter{\hbox{\includegraphics[width=0.19\textwidth]{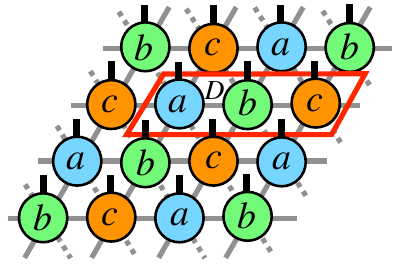}}} |\{s\}\rangle,
    \label{eq:ansatz}
\end{equation}
where $\{s\}$ represents spin configurations in the computational $S^z$ basis.  
Each site tensor has one physical index (black vertical line) labeling two states with spin-$1/2$ degree of freedom, and four auxiliary indices (grey solid lines) of bond dimension $D$. Pairs of neighboring site tensors are contracted through shared auxiliary indices, forming an effective square lattice. The nearest-neighbor exchange along one of the principal directions of the triangular lattice (gray dashed lines) becomes next-nearest neighbor on the effective square lattice. We evaluate the energy, order parameters, and other observables of $|\Psi(A)\rangle$ with the corner transfer matrix renormalization group algorithm~\cite{Nishino_1996, fishman18_ctmrg}.   
To find optimal tensors $A=[a,b,c]$ with increasingly large bond dimension $D$, amounting to $3\times 2D^4$ parameters, we perform gradient-based minimization of the energy using automatic differentiation~\cite{liao19-autodiff}. This bond dimension controls the accuracy of the iPEPS ansatz, setting an upper bound on the entanglement in the state, with $D=1$ corresponding to mean-field (product) states.

Given a ground state iPEPS wavefunction $\ket{\Psi_0(A)}$, the iPEPS excitation ansatz~\cite{laurens15-exci,laurens19-exci} is constructed by introducing a plane-wave superposition of a local defect wavefunction given by
\begin{equation}
    \ket{\Phi_\vk(B=[\tilde{a},\tilde{b},\tilde{c}])}=\sum_{\vx} e^{i\vk\cdot\vx} \ket{\Phi_\vx (B)}
    \label{eq:exciansatz}
\end{equation}
where $\ket{\Phi_\vx(B)}$ is obtained by replacing the tensor 
on site $\vx$ by an ``excited'' tensor from $B$. 
We note that the momentum $\vk$ is a good quantum number for the excited state $\ket{\Phi_\vk(B)}$ by construction. 
The $B$ tensors are chosen such that the excited state $\ket{\Phi_\vk(B)}$ is orthogonal to the ground state, $\langle\Psi(A)\ket{\Phi_\vk(B)}=0$.
In principle we can construct at most $M=3(2D^4-1)$ different excited tensors this way, here labeled $B_\alpha$, with $\alpha=1,\ldots,M$, for which, however, the 
 corresponding iPEPS excitations are, in general, not orthogonal, i.e., $\bra{\Phi_\vk(B_\alpha)}\Phi_\vk(B_\beta)\rangle\neq0$.
Under the assumption, that the low-lying eigenstates are supported on the subspace spanned by these iPEPS excitations $\{|\Phi_{\vk}(B_\alpha)\rangle\}$, we project the time-independent Schr\"odinger equation from the full Hilbert space onto this subspace, leading to a generalized eigenvalue equation for each momentum:
\begin{equation}
     \hat{H} P_{\mathbf{q}} |\psi\rangle = E_{\vq} P_\mathbf{q} |\psi\rangle,
\end{equation}
where $P_\vq=\sum_{\beta=1}^{M}\ket{\Phi_{\vq}(B_\beta)}\bra{\Phi_{\vq}(B_\beta)}$ 
is a projector on the iPEPS excitation manifold (also called tangent space).
In practice we keep only $n_b<M$ solutions, the excited eigenstates $\ket{\Phi_\vq(\tilde{B}_\alpha})$ with energies $\tilde{E}_{\vq\alpha}$, which are the most robust based on their norms, in order to resolve the gauge redundancy and filter unphysical states with very small norms~\cite{adpeps}. 
Now we can treat the dynamical spin structure factor which is defined as
\begin{align}
    S^{\sigma\sigma}(\vq,\omega)&\equiv\langle\Psi(A)|\hat{s}^\sigma_{-\vq}\hat{s}_{\vq}^\sigma\,\delta(\omega-\hat{H}+E_0)|\Psi(A)\rangle
    \label{eq:sqw}
\end{align}
where $\hat{s}^\sigma_{\pm\vq}=\sum_{\vx} \exp[\mp i\vq\cdot\vx]\hat{S}^{\sigma}_{\vx}$ are the $\sigma$ component ($\sigma=x,y,z$) of the spin operators in the momentum space representation and $E_0$ is the ground state energy.
By inserting the projector $\tilde{P}_\vq=\sum_{\alpha=1}^{n_b}\ket{\Phi_{\vq}(\tilde{B}_\alpha)}\bra{\Phi_{\vq}(\tilde{B}_\alpha)}$ constructed from the solutions  $\tilde{B}_\alpha$ into Eq.~\eqref{eq:sqw}, we obtain the final expression for the dynamical spin structure factor resolved on iPEPS excited states as a sum: 
\begin{align}
    S^{\sigma\sigma}(\vq,\omega)=\sum_{\alpha} p^\sigma_{\alpha}(\vq) \delta(\omega-\tilde{E}_{\vq\alpha}+E_0),
    \label{eq:sqw2}
\end{align}
where $p^\sigma_{\alpha}(\vq)= |\langle\Phi_\vq(\tilde{B}_\alpha)|\hat{s}^{\sigma}_{\vq} \ket{\Psi(A)}|^2$ is the corresponding spectral weight for each eigenstate. A Lorentzian broadening of the $\delta$-functions was used to model the experimental resolution. For further details, see SM.

\vspace{1mm}
\noindent
\textit{Field-induced phase transitions in} $\text{K}_2\text{Co}(\text{SeO}_3)_2$.--- First, in Fig.~\ref{fig:phase_diagram} we present the phase diagram for the compound obtained from iPEPS on the basis of the model in Eq.~\ref{eq:model}  and its magnetization curves.
The two supersolid phases, highlighted by yellow and green colors in Fig.~\ref{fig:phase_diagram}, exhibit both the non-zero in-plane transverse magnetization $m^{st}_{xy}$ (blue curve) indicating spontaneous U(1) symmetry breaking, and  
nonuniform $z$-direction magnetization $m_z^{st}$ (red curve) within the three-site unit cell, breaking the lattice $C_3$ symmetry.
Taken together, these are the two defining characteristics of a supersolid state~\footnote{Translating into the hard-core boson language, the two order parameters correspond to superfluid density $\langle b^\dagger\rangle$ and staggered occupation number $n_{st}=(n_A-n_B)/2$ (here, defined in terms of a two-sublattice model for simplicity), with $n_i=|\langle b_i^\dagger b_i\rangle|$ ($i\in\{A, B\}$)}.
When the magnetic field is increased, the two tilted spins in the SSY phase start canting towards the field direction, until the system enters the gapped up-up-down (UUD) phase characterized by the $m_z=1/3$ magnetization plateau.  As the magnetic field strength increases further, we find another supersolid phase before saturation, highlighted in green in Fig.~\ref{fig:phase_diagram}. 
Crucially, our iPEPS calculations suggest that this high-field SS state is the $\Psi$ ($\pi$-coplanar) state, in contrast to the ``V" state found in cluster mean-field study~\cite{yama14-xxz-cmf}, and it has not been investigated in detail in previous DMRG works~\cite{yama14-xxz-cmf,zhang06-dmrg_trgl_xxz}. We study its stability with $J_{xy}$ at $B_z=20$~T, corresponding to $g\mu_B B_{z}/J_{zz}=3$ in the DMRG phase diagram~\cite{zhang06-dmrg_trgl_xxz}. To distinguish the two states, the order parameter is chosen to be the minimal angle between mutual spins $\Delta\theta_{min}$, plotted in Fig.~\ref{fig:phase_diagram}(b) as a function of $J_{xy}/J_{zz}$. For the V state $\Delta\theta_{min}=0$ due to two collinear spins, while  $\Delta\theta_{min}>0$ for the $\Psi$ state.  We only obtain the V state for bond dimension $D=1$, consistent with the aforementioned cluster mean-field results~\cite{yama14-xxz-cmf}; however, for $D\geq 2$ our variational  correlated iPEPS strongly indicates that the V order yields to the $\Psi$ order for $J_{xy}/J_{zz}\leq0.4$, characterized by a finite $\Delta\theta_{min}$. 

\begin{figure}[thb]
    \centering
    \includegraphics[width=0.48\textwidth]{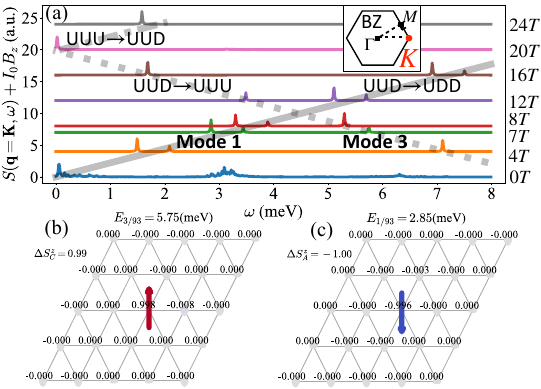}
    \caption{(a) Evolution of the dynamical spin structure factor $S(\vq, \omega)$ at the $K$ point under magnetic fields. The gray solid and dotted lines correspond to energy differences for different spin-flip (magnon) excitations at the mean-field level. For the supersolid phases at $B_z=0$ and $20$~T, we use $D=4$; for the UUD and FP states, we use $D=2$. (b) and (c) The real space visualization of the first and third lowest excited modes at the $K$ point, where $\langle S^z\rangle$ is evaluated on the $5\times 5$ patches with the excited tensor $B$ inserted at the center site and the ground state expectation values subtracted. The center in (b) is located in the $a$ sublattice (up spin), and the center in (c) is located in the $c$ sublattice (down spin). }
    \label{fig:evolution_excitation_k}
\end{figure}

Based on the optimized ground states $|\Psi(A)\rangle$, we use the excitation ansatz to compute the dynamical spin structure factor $S(\vq,\omega)$ as per Eq.~\eqref{eq:sqw2}. In Fig.~\ref{fig:evolution_excitation_k}(a), we plot the structure factors at the $K$ point (corner of the Brillouin zone) under different magnetic fields. As the field strength increases, the system enters the UUD phase, where structure factors feature finite energy gaps, further corroborating  its gapped nature. As the magnetic field $B_z$ keeps increasing and becomes larger than $10$~T, the energy gap starts to decrease and closes when entering the SSP phase around $B_z=20$~T. Eventually, when the field $B_z$ is strong enough, the system becomes fully polarized (FP).

Looking further into the evolution of the spectrum, we note that there are always three well-defined (quasiparticle-like) finite-energy modes in the UUD phase. It is natural for one to make connections between these modes and spin-flip (magnon) excitations given the simple product-state-like picture of the state. 
To confirm this, we compute the mean-field energy difference between the UUD state and the up-down-down (UDD) state as a function of the magnetic field $B_z$, plotted as a gray solid line in Fig.~\ref{fig:evolution_excitation_k}(a). We note that this mean-field line always crosses the center of the two right-propagating modes in the excitation spectrum in the UUD phase. Similarly, we compute the mean-field energy difference between the UUD and FP (up-up-up) states as a function of field,  shown by the gray dashed and dash-dotted lines in Fig.~\ref{fig:evolution_excitation_k}(a). Again, this energy difference matches perfectly with the left-propagating mode in the spectrum. 
Through this analysis, we unambiguously identify the physical origin of these excitation modes in the UUD phase.

As a direct probe, we further present a real-space visualization~\cite{Ponsioen20-exci_peps} for the aforementioned modes at $B_z=7$~T, obtained by considering the real-space effect of an excited tensor $B$ in one term of the excitation ansatz in Eq.~\eqref{eq:exciansatz}.
Mode 1 [the peak immediately to the left
of the solid gray line in Fig.~\ref{fig:evolution_excitation_k}(a)] corresponds to the up-to-down spin flip, as shown in Fig.~\ref{fig:evolution_excitation_k}(b), since the excited tensor results in a change in the local $S^z$ expectation value ($\Delta S^z \simeq -1$). We note that the twofold degeneracy of the excitation is lifted by the $J_{xy}$ term, resulting in a splitting $\Delta E_{1,2}=\pm 3J_{xy}/2$ in the excitation energy per triangle. Mode 3 (dotted gray line), by contrast, has a positive change in the on-site $S_z$ expectation value $\Delta S^z \simeq +1$, which corresponds to the down-to-up spin flip, as shown in Fig.~\ref{fig:evolution_excitation_k}(c). As before, there is a $-3J_{xy}/2$ energy shift due to the hopping term that introduces dynamics to the static defect.

\begin{figure}[tb]
    \centering
    \includegraphics[width=0.5\textwidth]{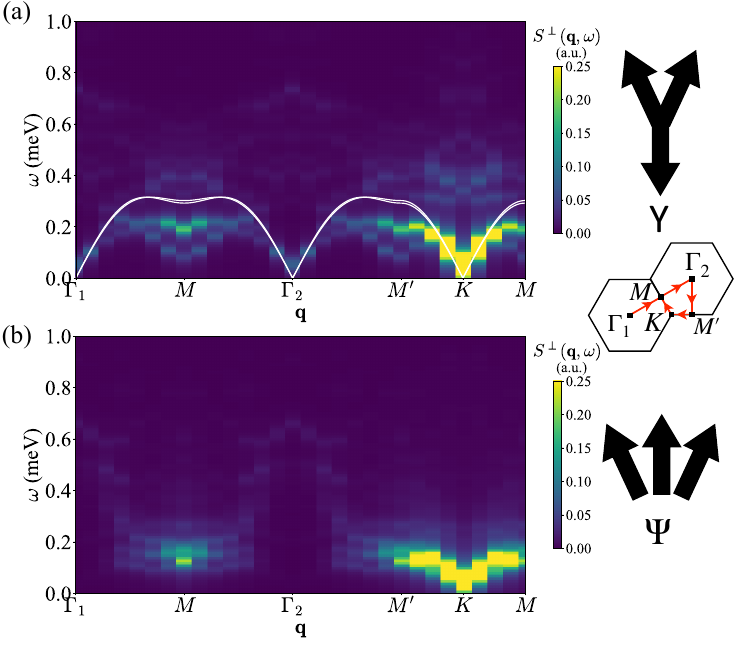}
    \caption{Transverse spin excitation spectrum [$S^\perp(\vq,\omega)=S^{xx}(\vq,\omega)+S^{yy}(\vq,\omega)$] showing only the lowest branch for (a) the Y phase at $B_z=0$ and (b) the \yi{$\Psi$} phase at $B_z=20$~T along the momentum path, $\Gamma_1$-$M$-$\Gamma_2$-$M'$-$K$-$M$, as indicated by the red arrows on the right. We use $D=3$ and keep $n_b=50$ excited states for the lowest branches in the two supersolid phases. The white lines in the top panel are obtained with linear spin wave theory. The Lorentzian broadening factor is chosen to be $\eta=0.02$~meV, slightly lower than the experimental resolution $\eta=0.05$~meV, allowing us to resolve different branches at the $M$ point in the SSY phase. \vspace{-3mm}
    }
    \label{fig:sqw_low_energy}
\end{figure}

\vspace{1mm}
\noindent
\textit{Low-energy spectrum in the supersolid phases.---} Unlike the UUD and FP phases which have well-defined quasi-particle excited modes, the two supersolid phases spontaneously break the global U(1) symmetry, which gives rise to the gapless Goldstone modes. Here, we present the low-energy transverse spin excitation spectrum for the two supersolid phases, as shown in Fig.~\ref{fig:sqw_low_energy}. We note that the transverse dynamical structure factor, $S^\perp=S^{xx}+S^{yy}$ has both in-plane and out-of-plane components (in-plane refers to the direction where the magnetic order forms). 
For the zero-field state, we observe the Goldstone mode at the $K$ point, 
suggesting the three-sublattice order of the SSY supersolid phase. 
Our results also show the strong downward renormalization at the $M$ point \yi{related} to the roton-like excitations~\cite{zheng06-roton, chen19} in the dispersion compared to linear spin wave theory.
The zero-field data we obtain are in good agreement with the experimental data of Ref.~\cite{chen2024phase} (see the SM for a detailed comparison).

In contrast to the low-field SSY phase, however, the analysis is more complicated for the high-field supersolid phase, where, as we remarked earlier, the V order obtained at the mean-field level ($D=1$) yields to the $\Psi$ order for higher bond dimensions $D\geq2$. 
Thus, the linear spin wave theory based on the V state does not provide a valid starting point.
Still, we find the smallest gap again at the $K$ point [see Fig.~\ref{fig:sqw_low_energy}(b)], similar to the SSY phase, indicating a Goldstone mode associated with the formation of a three-sublattice order.
The dispersion around the $M$ point is decidedly not roton-like and is instead almost flat.
We anticipate that future INS studies will identify these key signatures in the high-field supersolid phase.

\begin{figure}[tb]
    \centering
    \includegraphics[width=0.49\textwidth]{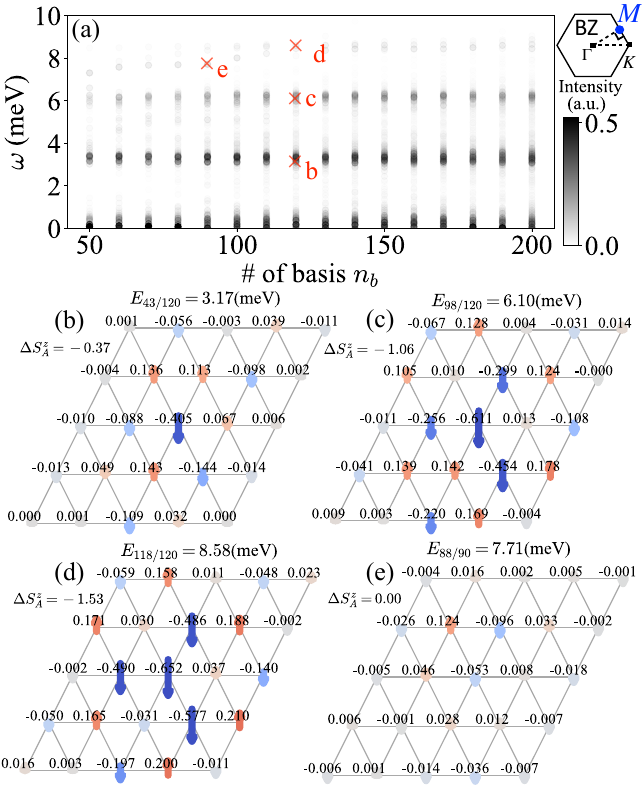}
    \caption{(a) Basis dependence of the dynamical spin structure factors $S(\vq,\omega)$ for the zero-field state with $D=3$ at the $M$ point. The intensity depends on both the norm of the individual states (colors) and the density of states 
    (opacity). The lowest three ``quantized" continua are robust with different number of excited states kept, while the fourth one ($\omega_3=3J_{zz}$) will drop with less than $n_b=120$ excited states. (b)--(e) Real space visualization of the excited states at the $M$ point at different energies which are picked from each continuum: (b) $\omega_1=J_{zz}$, $n_b=120$; (c) $\omega_2=2J_{zz}$, $n_b=120$; (d) $\omega_3=3J_{zz}$, $n_b=120$; and (e) $\omega_3=3J_{zz}$, $n_b=90$. All configurations are centered at the $a$ sublattice.\vspace{-3mm}}
    \label{fig:basis_dependence_M}
\end{figure}

\noindent
\textit{Quantized continua in the SSY phase.---} We now turn our attention back to the low-field SSY phase. 
Besides the low-energy spectrum shown earlier in Fig.~\ref{fig:sqw_low_energy}(a), we find a peculiar set of nearly dispersionless features in our simulations at higher energies (see the SM for the complete excitation spectrum). 
These features are most easily seen on the Brillouin zone boundary, shown  in Fig.~\ref{fig:basis_dependence_M}, where we plot $S(\vq=M,\omega)$ against the increasing cutoff on the number of excited states $n_b$ kept in the simulation. 
Apart from confirming that our results are well converged with respect to this cutoff for $n_b\gtrsim 120$, Fig.~4 allows us to nicely illustrate the three sets of excitations located at energies $\omega_1\simeq3$~meV, $\omega_2\simeq6$~meV and (fainter) $\omega_3\simeq9$~meV.
What does the apparent energy quantization signify and what is the physical explanation for the corresponding eigenstates?
Our excited PEPS method 
provides suggestive evidence
that at least the quantized feature at $\omega_2$ corresponds to the two-magnon excitation. This is best illustrated in the real-space distribution of $S^z$ expectation values, as previously done for the UUD phase in Fig.~\ref{fig:evolution_excitation_k}, now shown for the SSY phase 
in Figs.~\ref{fig:basis_dependence_M}(b)--\ref{fig:basis_dependence_M}(d).
First, we observe that the eigenmode at energy $\omega_1=J_{zz}\simeq3$~meV, shown in Fig.~\ref{fig:basis_dependence_M}(b), corresponds to a one-magnon (down-spin-flip) state.
Note, however, that although the excited tensor $B$ is placed only at the center of the supercell, it should not be thought of as always representing a single spin-flip -- our data show unequivocally that insertion of a single $B$ can affect multiple spins in its neighborhood, with the distance proportional to the correlation length in the ground state $|\Psi(A)\rangle$ on which the excitation is built. Here, at $B_z=0$ these correlation lengths are $\xi(D=3)=1.1$ and $\xi(D=4)=1.4$.
Similarly, we find that the spectral feature at $\omega_2\simeq 2J_{zz}$ corresponds to 
a two-spin-flip (i.e., two-magnon) excitation (consistent with Ref.~\cite{Xie2023}), 
as seen from Fig.~\ref{fig:basis_dependence_M}(c). \yi{Furthermore, we note that only the transverse components of the structure factor contribute to those $n\geq1$ branches, corroborating their connection to spin flips. 
}

\vspace{1mm}
\noindent
\textit{Discussion.---} 
In this work, we employed iPEPS tensor networks for both the ground and excited states to investigate
the two-dimensional Ising-like triangular lattice compound $\text{K}_2\text{Co}(\text{SeO}_3)_2$ in an applied magnetic field, with a particular focus on the field-induced supersolid phases. 
Our numerical simulations for the strong easy-axis XXZ model on the triangular lattice proposed to describe the compound, corroborate the existence of the two supersolid phases under both low (SSY) and high (SSP) magnetic fields. The excitation spectrum we obtained for the zero-field SSY phase lends itself to interpretation as the U(1) Goldstone mode associated with the spin superfluidity and matches the experimental data very nicely, thus providing important numerical evidence of the existence of supersolids in this material. 
Interestingly, at zero-field, the scaling of the superfluid order parameter to $D\to\infty$ suggests that it either vanishes or remains very weak, a result also noted recently in Ref.~\cite{ulaga2024easyaxisheisenbergmodeltriangular}. The same applies to the gap at the $K$ point (see the SM). This observation, together with signatures of the continuum emerging  directly from the $K$ point in the INS data~\cite{zhu2024continuum} underscores the need for further investigation of the zero-field case.
The analysis of our excited iPEPS states at higher energies gives insight into the multi-spin-flip nature of the excitation continua $\omega_n = nJ_{zz}$ and their apparent energy quantization, which has been observed experimentally in the SSY phase.
In addition, our simulated spectra for the high-field SSP phase provide concrete predictions, which should be verified in future INS experiments.

Overall, our work demonstrates the power of iPEPS as a comprehensive framework to describe ground and excited state properties of frustrated magnets. This should further motivate its future applications, in particular to challenging cases of systems which might host quantum spin liquids
such as triangular lattice compounds with strong beyond-nearest-neighbor interactions~\cite{xu2023realization,gong19-gs_dmrg_j1j2j3_trgl}. Similarly, the interplay between magnetic field and frustration can give rise to exotic liquid-like states with algebraically decaying spin correlations~\cite{sur24-field-induced},
offering another promising direction to explore due to the tunability of magnetic fields in the experiments.


\vspace{1mm}
\begin{acknowledgments}
\noindent
\textit{Acknowledgements}.---
We would like to thank T. Chen for the early communication on the experimental neutron scattering findings for $\text{K}_2\text{Co}(\text{SeO}_3)_2$. We thank L. Chen, Z. Wang, P. Corboz and J.-Y. Chen for fruitful discussions.
The ground states were simulated using the \textit{peps-torch}~\cite{pepstorch} package and excited state calculations were performed using the \textit{ad-peps} package~\cite{adpeps} developed by B.P. and extended to triangular lattices by Y.X.
The tensor network calculations by Y.X. and A.H.N. were supported by the U.S. Department of Energy under
the Basic Energy Sciences award No. DE-SC0025047.
J.H. and B.P. acknowledge support from the European Research Council (ERC) under the European Union's Horizon 2020 research and innovation programme (Grant Agreement No. 101001604). J.H. acknowledges support from the Swiss National Science Foundation through a Consolidator Grant (iTQC, TMCG-2\_213805).
The computing resources at Rice University were supported in part by the Big-Data Private-Cloud Research Cyberinfrastructure MRI-award funded by the U.S. National Science Foundation under grant CNS-1338099 and by Rice University's Center for Research Computing (CRC).

\vspace{1mm}
\noindent
\textit{Note added}. In the process of preparing the manuscript, we became aware of several recent works~\cite{sheng2024continuum,chi2024dynamical,gao2024spin} which use DMRG as well as iPEPS  (yet with a different choice of a unit cell) to study the supersolid phases in another easy-axis anisotropic triangular lattice antiferromagnet $\text{Na}_2\text{BaCo}(\text{PO}_4)_2$. Our results for $\text{K}_2\text{Co}(\text{SeO}_3)_2$ are consistent with their findings in the low-energy dispersion given that these two materials both correspond to the SSY phase in zero field. By contrast, due to the relatively small anisotropy $J_{zz}/J_{xy}$ in $\text{Na}_2\text{BaCo}(\text{PO}_4)_2$, their excitation spectrum for the low-field SSY phase does not have the quantized ($\omega_n=nJ_{zz}$) continua discussed in this work.
\end{acknowledgments}

\UseRawInputEncoding
\renewcommand{\emph}[1]{\textit{#1}}
\bibliography{ref_exci_kcso}

\clearpage
\setcounter{equation}{0}
\setcounter{figure}{0}
\setcounter{table}{0}
\makeatletter
\renewcommand{\theequation}{S\arabic{equation}}
\renewcommand{\thefigure}{S\arabic{figure}}
\renewcommand{\bibnumfmt}[1]{[#1]}
\renewcommand{\citenumfont}[1]{#1}


\begin{widetext}
\begin{center}
	\Large{Supplementary Materials for ``\thistitle"}
\end{center}
\end{widetext}

\section{Comparison with experimental data at $B_z=0$\label{sup:comparison}}
\begin{figure}[tbh]
    \centering
    \includegraphics[width=\textwidth]{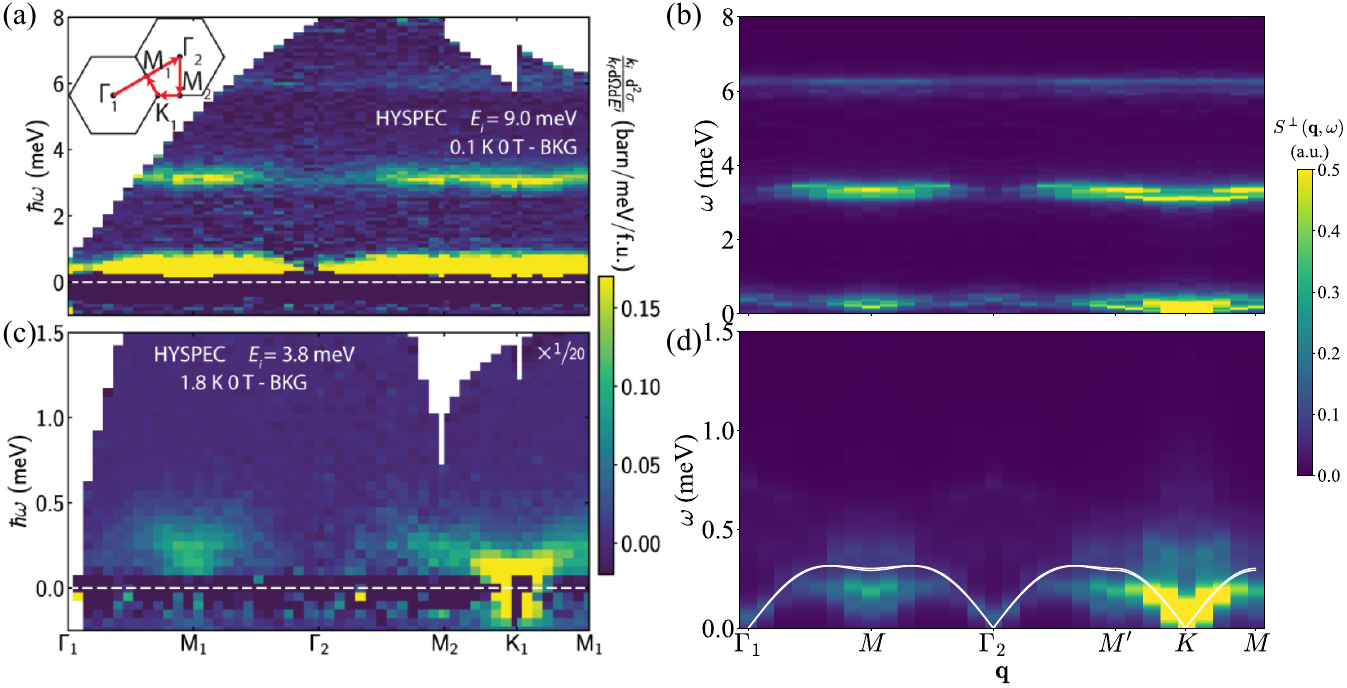}
    \caption{Comparison between (a) and (c) the experimental neutron scattering data from Ref.~\cite{chen2024phase} (with permission from the corresponding author) and (b,d) our $D=3$ simulation results. The top two panels are plotted in a large energy scale. The Lorentzian broadening factor is chosen to be $\eta=0.05$~meV, which is the same as the experimental resolution. We keep $n_b=120$ for (b) and $n_b=50$ for (d).}
    \label{fig:comparison_kcso}
\end{figure}

\vspace{-5mm}
We compare side-by-side the excitation spectra obtained by iPEPS excitation ansatz with those from the inelastic neutron scattering (INS) experiments~\cite{chen2024phase}.
As panels~\ref{fig:comparison_kcso}(a) and (b) demonstrate, there is a clear one-to-one correspondence between the nearly flat branches of excitations at energies $\omega=nJ_{zz}\ (n=0,1,2)$ we obtained and the INS data. Zooming in on the energy window below 1.5~meV, the dispersion of the lowest branch  in Fig.~\ref{fig:comparison_kcso}(d) is in a very good agreement with the experimental data in panel (c).

Please note that due to the basis dependence shown in Fig.~4 in the main text, a large enough number of considered excited states is needed to recover the complete spectrum including the aforementioned branches $\omega=nJ_{zz}$. However, increasing the number $n_b$ of the excited states kept in the simulation might result in a loss of local resolution since there may be ``false" excited states -- solutions with small norms entering the effective Hamiltonian matrix, causing numerical problems for the diagonalization. Hence, we choose to keep different numbers of excited states for the two different energy regimes: $n_b=120$ in panel (b) and $n_b=50$ in panel (d) of Fig.~\ref{fig:comparison_kcso}.
\onecolumngrid
\twocolumngrid

\begin{figure}[tbh]
    \centering
    \includegraphics[width=0.48\textwidth]{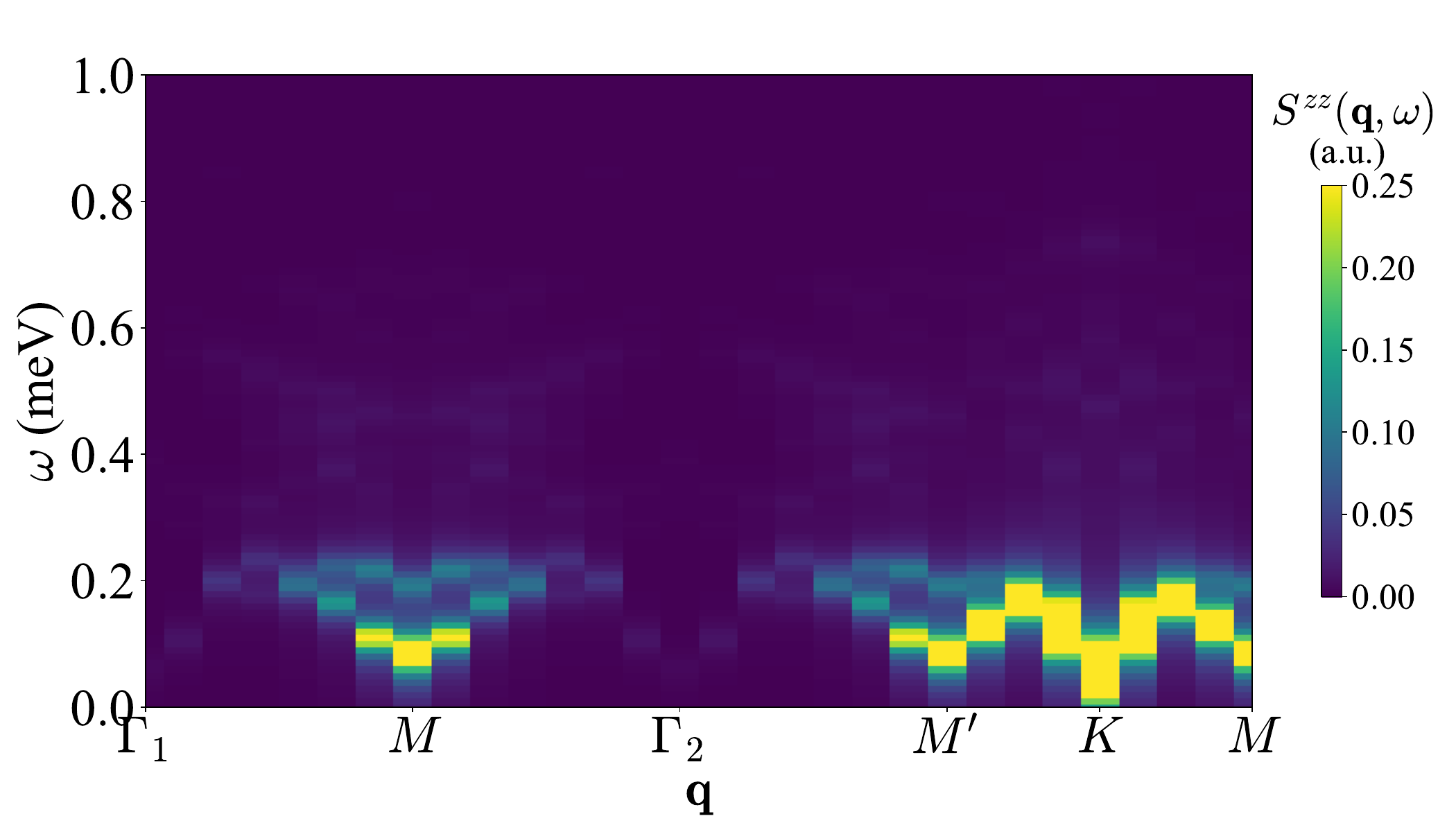}
    \caption{Out-of-plane component of the dynamical spin structure factor, $S^{zz}(\boldsymbol{q},\omega)$, for the Y phase at $B_z=0$, produced by $D=3$ PEPS simulations with $n_b=50$ states kept. The Lorentzian broadening factor is chosen to be $\eta=0.02$~meV.}
    \label{fig:szz_lb}
\end{figure}

Note that in the above comparison, we choose $S^\perp$ in our simulations to compare with the experimental data, because those higher branches do not have significant contributions from $S^{zz}$. In the neutron scattering experiment, however, the scattering intensity is a superposition of $S^{\perp}$ and $S^{zz}$~\cite{chen2024phase}. In fact, it is actually $S^{zz}$ component that dominates the overall scattering intensity due to the large magnetization in $z$-direction. Hence, here we also present the $S^{zz}$ component of, as shown in Fig.~\ref{fig:szz_lb}. It can be seen that the lower roton-like minimum, which is located at $E\approx0.1$~meV, is much more intensive than the higher one, which agrees better with the experimental scattering spectrum.

\section{Estimate of Pseudo-Goldstone Gap}
\begin{figure}[tbh]
    \centering
    \includegraphics[width=0.48\textwidth]{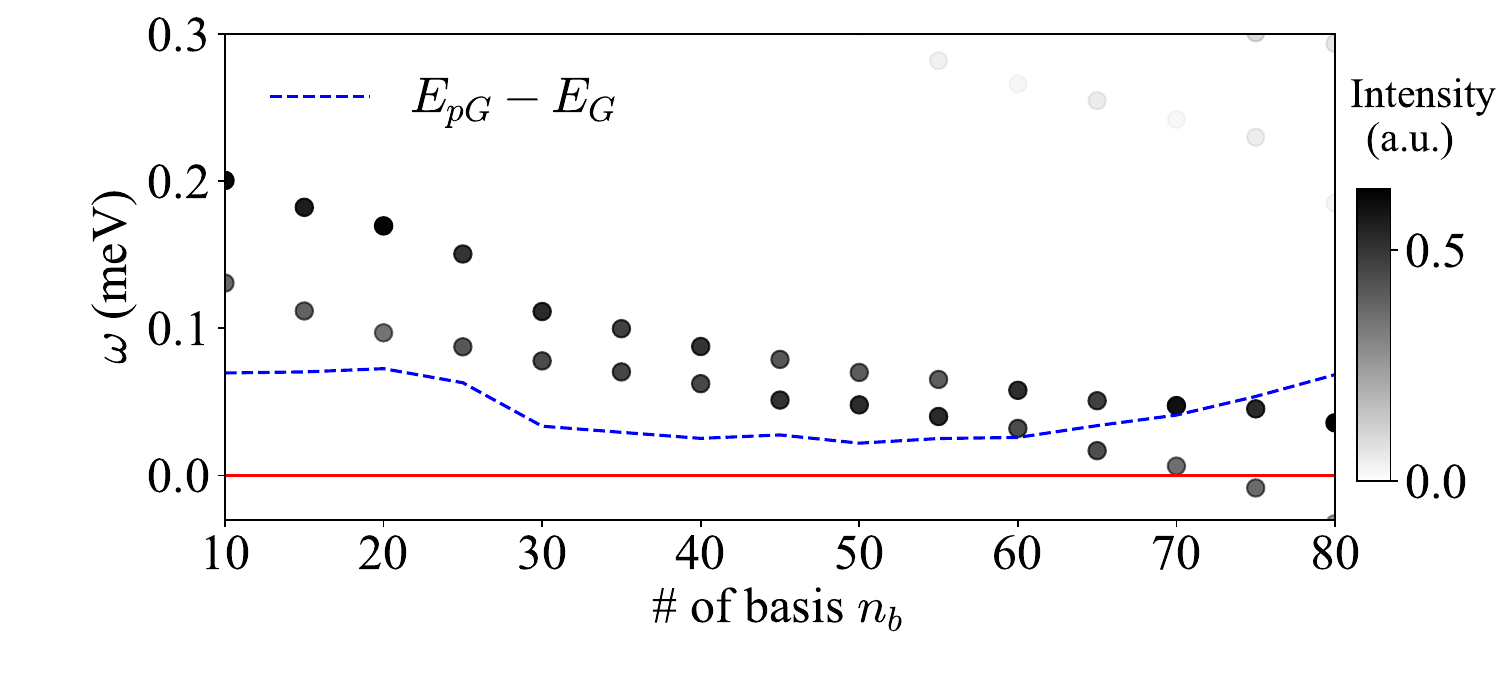}
    \caption{The basis dependence of low-lying excited modes at $K$ point, for $D=3$ PEPS simulations.}
    \label{fig:pGS_basis_dep}
\end{figure}

\vspace{-3mm}
The excitation ansatz for iPEPS constructed the tangent space of the ground state using 
quasi-particle-like excitations. Such construction allows us to have access to individual excited modes which are broadened afterwards to mimic the finite-temperature behavior. This caveat of the method instead makes it easier to resolve different modes that have very close energies. Here, we demonstrate this by looking at the pseudo-Goldstone mode that is found in the latest neutron scattering experiments~\cite{chen2024phase,zhu2024wannierstatesspinsupersolid}. In Fig.~\ref{fig:pGS_basis_dep}, we plot the basis-size dependence of the low-lying excited modes at the K point. The two dominant modes correspond to the Goldstone (mainly from $S^\perp$) and pseudo-Goldstone (mainly from $S^{zz}$) modes. Note that the finite energy gap of the Goldstone mode is due to the finite bond dimension effect. We shall also point out the numerical instability regarding the basis-size dependence and the negative energy modes ($n_b\geq75$), which was first pointed out in Ref.~\cite{adpeps}, and can be in principle improved by including the projector derivatives~\cite{adpepsv3}, or using the coarse-grained ansatz used in Ref.~\cite{chi22-exci_xxz_trgl}.

Despite the limitations mentioned above, we can still choose a certain range that is relatively stable. To determine that, we choose compute the difference between the pseudo-Goldstone and Goldstone mode for different numbers of states kept. And within that range, we choose the largest $n_b$ for which the lowest Goldstone mode is dominant. In this case, we choose $n_b=55$. and the corresponding estimated pseudo-Goldstone gap is $E_{pG}=0.06525$~meV, which agrees quite well with the experimental results ($E_{pG}^{\text{exp}}\approx60\mu$eV)~\cite{chen2024phase,zhu2024wannierstatesspinsupersolid}. Finally, we point out that the pseudo-Goldstone mode is more intensive than the Goldstone mode in the $S^{zz}$ component, suggesting its being connected to the lower roton branch.

\section{Low field ``Y" phase at $B_z=0.5$T\label{sup:low_field_Y}}
\begin{figure}[thb]
    \centering
    \includegraphics[width=0.49\textwidth]{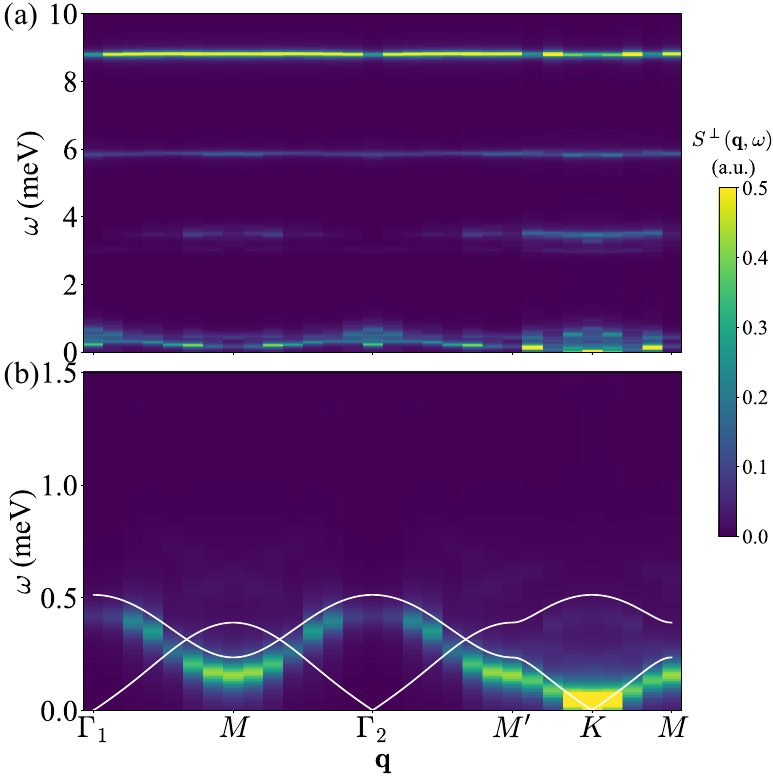}
    \caption{Transverse spin excitation spectrum for the Y state at $B_z=0.5$~T. The white lines correspond to the linear spin wave theory. We use $D=3$ here and keep $n_b=120$ for (a) and $n_b=50$ for (b). The Lorentzian broadening factor is chosen to be $\eta=0.02$~meV.}
    \label{fig:sqw_hz_0.22575}
\end{figure}

\vspace{-3mm}
In this section, we present the excitation spectrum obtained by the iPEPS excitation ansatz for the low-field ($B=0.5$~T) SSY phase, shown in Fig.~\ref{fig:sqw_hz_0.22575}. As argued in the main text, the excitation spectrum of this low-field SSY state exhibits features of superfluidity and solidity -- the gapless Goldstone mode at the $K$ point and the high-lying quantized branches. 
The presence of the external magnetic field suppresses quantum fluctuation, resulting in more well-defined low-energy branches as shown in Fig.~\ref{fig:sqw_hz_0.22575}(b), compared to the more diffuse excitations in the zero-field state shown in Fig.~3a in the main text.

\section{UUD and FP phases\label{sup:uud_fp}}
\begin{figure}[thb]
    \centering
    \includegraphics[width=0.45\textwidth]{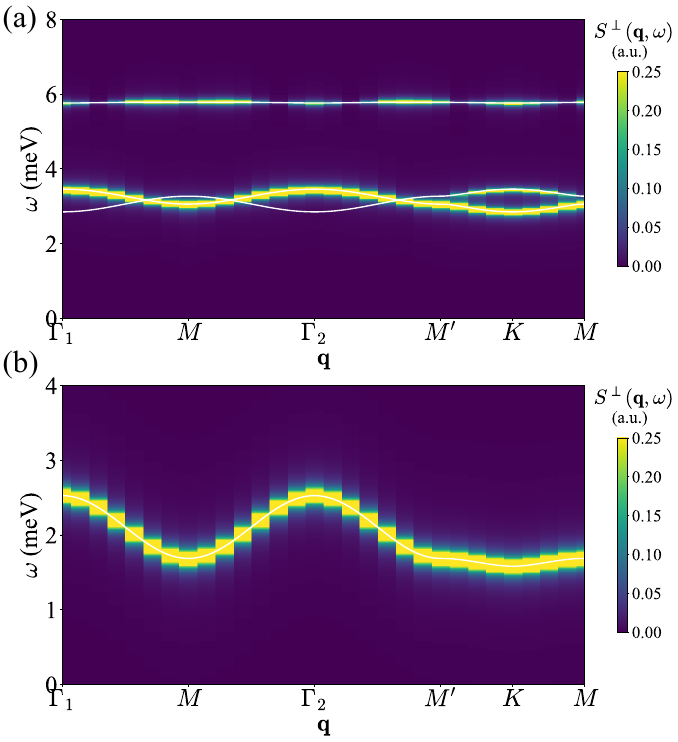}
    \caption{Transverse spin excitation spectrum for (a) the UUD phase at $B_z=7$~T and (b) the FP state at $B_z=24$~T. The white lines correspond to the linear spin wave theory.}
    \label{fig:sqw_uud_fp}
\end{figure}

\vspace{-4mm}
In this section, we present the excitation spectrum obtained by the iPEPS excitation ansatz for the UUD and the FP phases. As  Fig.~\ref{fig:sqw_uud_fp} demonstrates, the well-defined quasi-particle-like modes in the UUD and FP phases agree perfectly with those computed by the linear spin wave theory. This is to be expected, since the strong magnetic field suppresses the transverse field fluctuations in these phases. However, it is worth mentioning that in Fig.~\ref{fig:sqw_uud_fp}(a), one of the branches for the $UUD$ state at $B_z=7$~T has vanishing intensities along the $\Gamma_1-M-\Gamma_2-M'$ path. This is due to the fact that the momentum along any $\Gamma-M$ path is always parallel to one of the three lattice vectors on the triangular lattice and hence cannot lift the degeneracy.

\subsection{Mean-field analysis for spin-flip excitation}
In Fig.~2(a) of the main text, we plot the energy gap of spin-flip modes (gray lines), at the mean-field level. They are computed by comparing the mean-field energies for different configurations, including the up-up-down (UUD), up-down-down (UDD) and up-up-up (UUU or FP) patterns. Here, we give the expressions for the mean-field energies.  
\begin{align}
    E_{UUD} &= -\frac{3}{4}J_{zz} - \frac{1}{2} g\mu_B B_z \\
    E_{UDD} &= -\frac{3}{4}J_{zz} + \frac{1}{2} g\mu_B B_z\\
    E_{UUU} &= \frac{9}{4}J_{zz} - \frac{3}{2} g\mu_B B_z
\end{align}

\section{Finite-$D$ scaling\label{sup:scaling}}
\subsection{Scaling for ground state energies}

Here, we analyze the finite bond dimension effects for the ground states at $B_z=0, 0.5$~T and $20$~T. We extract the scaling of ground state energy with respect to the bond dimension in the form $E_{gs}(D) = E_{gs}(\infty) + \alpha/D^\eta$, fitting the unknown exponent $\eta$ as well as the coefficient $\alpha$ to extract the extrapolated value of energy as $D\to \infty$. The resulting fitting parameters are shown in the legend of Fig.~\ref{fig:gs_scaling}(a,c,e).

We note that in our simulations the environment bond dimension $\chi$, which governs the precision of CTMRG method, is chosen to be sufficiently large ($\chi\geq5D^2$) to converge ground state energy as well as the order parameters with respect to $\chi$. For both the correlated SSY and SSP phase, we note that a $D=3$ iPEPS can already give a very good approximation of the ground state energy, which has a difference of only about $0.001$~meV (per site) to the extrapolated infinite-$D$ value for the zero-field SSY state (and the difference is even smaller for the Y state at $B_z=0.5$~T and the $\Psi$ state at $B_z=20$~T).

Another way to present the data is by studying the scaling of energies with respect to the correlation length $\xi$ extracted from the transfer matrices. This has been shown to be a more precise approach to extrapolate the energies and order parameters in the spontaneous symmetry breaking states~\cite{corboz18-finite-correlation-scaling,hasik21-scaling-neel,lauchli18-finite-correlation-length-scaling}. We use the scaling relation $E_{gz} = E_{gs}(\infty) + \beta_3/\xi^3 + \beta_4/\xi^4$ to leading order in powers of $1/\xi$~\cite{lauchli18-finite-correlation-length-scaling}. The results are shown with the dashed lines in Fig.~\ref{fig:gs_scaling}(a,c,e) and give very similar extrapolations to those obtained earlier with finite-D scaling.

\subsection{Scaling for the order parameters}

\begin{figure}[thb]
    \flushleft
    \includegraphics[width=0.49\textwidth]{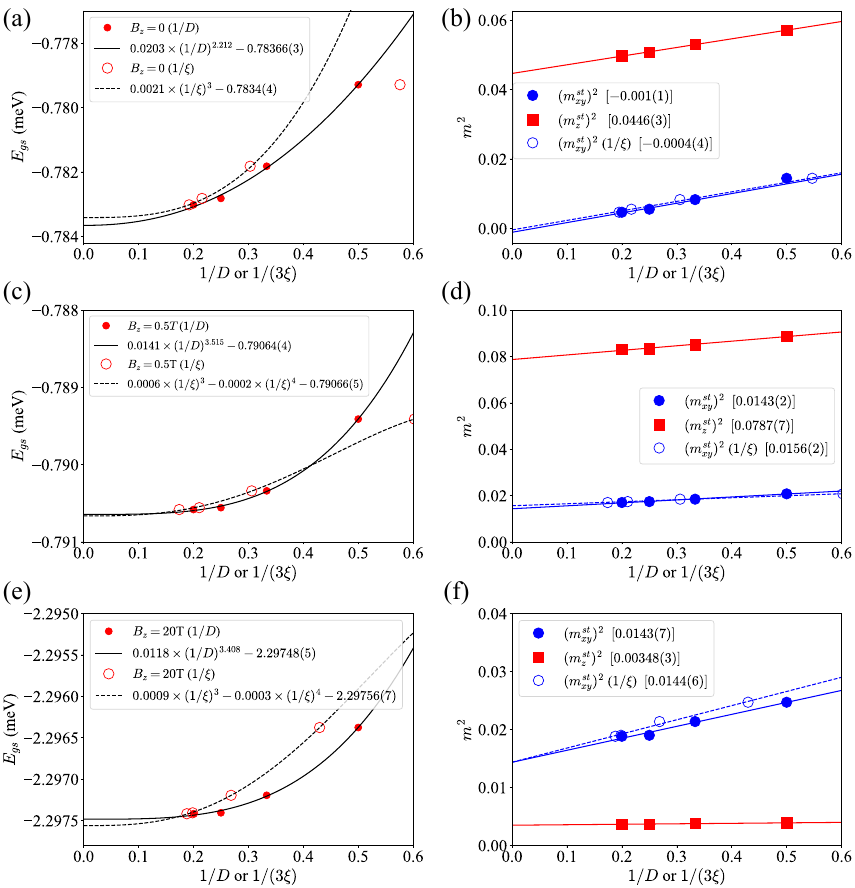}
    \caption{Finite bond dimension and correlation length scaling for the supersolid ``Y" states at $B_z=0$ and $B_z=0.5$~T, and the supersolid ``$\Psi$" state $B_z=20$~T -- (a,c,e) the ground state energy $E_{gs}$ and (b,d,f) the staggered transverse magnetization $m^{st}_{xy}$, the staggered and uniform out-of-plane magnetization $m_z^{st}$ and $m_z$ versus inverse bond dimension $1/D$ (solid markers and solid lines) or inverse correlation length $1/\xi$ (empty markers and dashed lines). In (b,d,f), the values in the brackets correspond to the extrapolated values for different order parameters. }
    \label{fig:gs_scaling}
\end{figure}

We now turn to the scaling of the order parameters (measured in the ground state) as functions of the bond dimension $D$ and correlations length $\xi$. The scaling formula for the U(1) symmetry breaking order parameter -- the staggered transverse magnetization $m_{xy}^{st}$ -- is given by $[m_{xy}^{st}(\xi)]^2=[m_{xy}^{st}(\infty)]^2+a/\xi +O(1/\xi^2)$~\cite{lauchli18-finite-correlation-length-scaling}. We note that there is a spurious symmetry breaking pattern for the $D=2$ state -- the correlation lengths in $x$ and $y$ directions have a relative difference more than $10\%$ ($\xi_x=0.6410, \xi_y=0.5786$). Hence, we ignore the $D=2$ state when fitting, as shown in Fig.~\ref{fig:gs_scaling}(a,b).

The scaling indicates that for both small, $B_z=0.5$~T, and high fields, $B_z=20$~T, the superfluid order parameter will remain  finite with only modest corrections. Instead, for the zero-field state, both the finite-$D$ and the finite correlation length scaling of the superfluid order parameter $(m_{xy}^{st})^2$ yield almost vanishing extrapolated values, as shown by Fig.~\ref{fig:gs_scaling}(b). Hence, the possibility of the U(1) symmetry being restored  cannot be excluded. This finding is consistent with the result in a recent numerical study using finite-temperature Lanczos method~\cite{ulaga2024easyaxisheisenbergmodeltriangular}.

In our iPEPS simulations, the spectral features were obtained from finite-$D$ iPEPS and they agree very well with the experimental data taken at small and zero fields. In these states, the U(1) symmetry is spontaneously broken. If the superfluid order parameter truly vanishes at $B_z=0$, as the above scaling analysis indicates, this would  mean that the U(1) symmetry is restored and the observed excitation spectra cannot be simply interpreted through the lens of spontaneous symmetry breaking and associated Goldstone mode. This scenario is likely specific to zero-field case, since a small finite field $B_z>0$ stabilizes the superfluid, which can be seen from the scaling of $m^{st}_{xy}$ in Fig.~\ref{fig:gs_scaling}(d) and from the appearance of a clear gapless mode at $K$ point in Fig.~\ref{fig:sqw_hz_0.22575}(b).

\subsection{Scaling for the excited states}

\begin{figure}[tbh]
    \centering
    \includegraphics[width=0.9\linewidth]{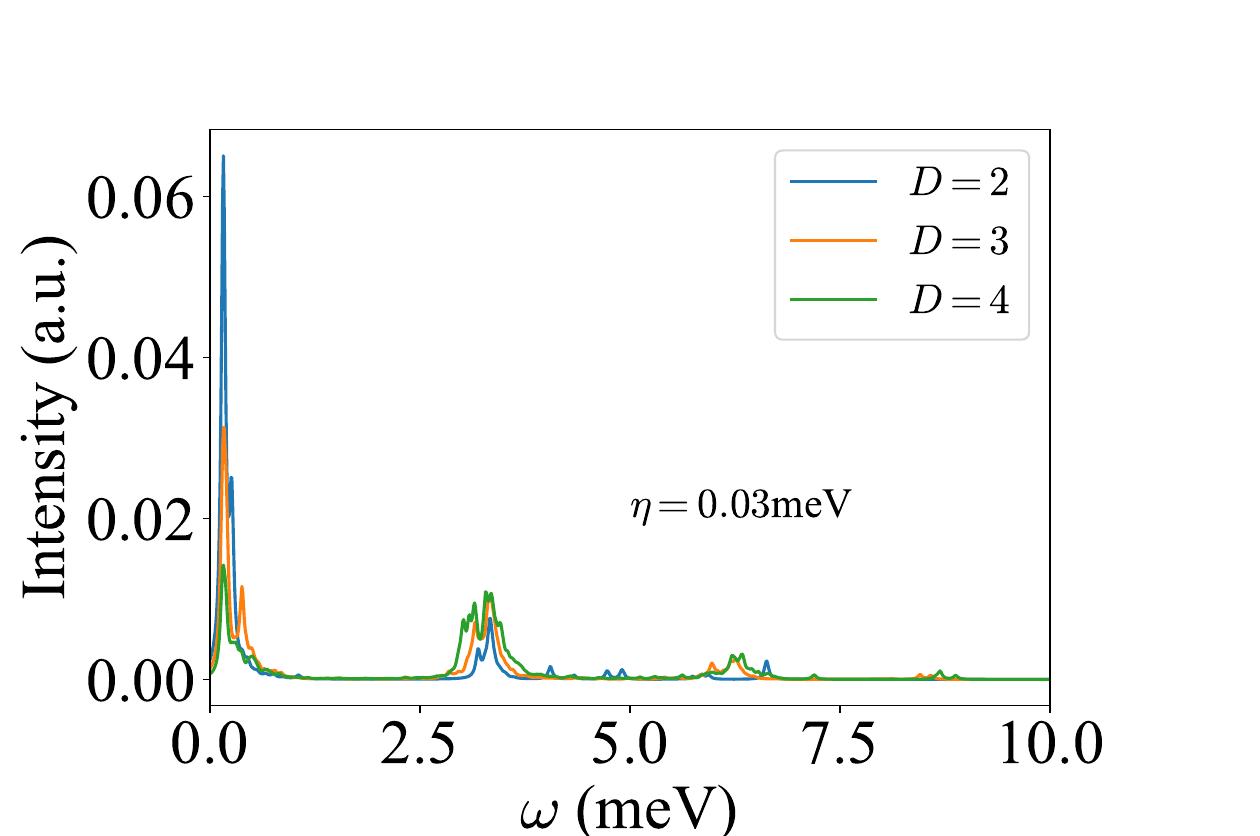}
    \caption{Total dynamical spin structure factor $S^{tot}(\boldsymbol{q}, \omega)=S^{xx}+S^{yy}+S^{zz}$ at the $M$ point for $D=2, 3, 4$. The numbers of excited states kept are $40, 120, 180$ for $D=2, 3, 4$ respectively, which are chosen to be the most robust against change of basis number. The Lorentzian broadening factor is chosen to be $\eta=0.03$~meV.}
    \label{fig:sup:exci_scaling_m_point}
\end{figure}

In this section, we show how the excitation spectrum at the $M$ point changes as we increase the bond dimension, demonstrated by Fig.~\ref{fig:sup:exci_scaling_m_point}. First, the  quantized features at energies $\omega_n=nJ_{zz}$ are partially present for $D=2$ and completely recovered at higher bond dimensions $D=3, 4$. We keep the same number of excited states $n_b=60$ throughout (see Fig. 4 in the main text for $n_b$ dependence). Then, we further look into the three different branches separately. For the lowest branch $\omega_0$, there is always a sharp peak ($\omega_0=0.119$~meV, $0.125$~meV, $0.106$~meV for $D=2,3,4$, respectively), followed by a smaller peak or a bump that decays gradually. 
The $\omega_1=J_{zz}$ and $\omega_2=2J_{zz}$ branches  can be clearly observed for $D>2$ state, with $\omega_1$ branch showings signs of two broad peaks. The weakest $\omega_3=J_{zz}$ branch only starts to develop for $D>2$.
We note that, as the bond dimension increases, the intensities of different branches will re-distribute. In particular, the relative intensity of the $\omega_0$ peak will drop, as Fig.~\ref{fig:sup:exci_scaling_m_point} shows. 

\section{Methodological advantages compared with coarse-grained ansatz}

\begin{figure}[tbh]
    \centering
    \includegraphics[width=0.9\linewidth]{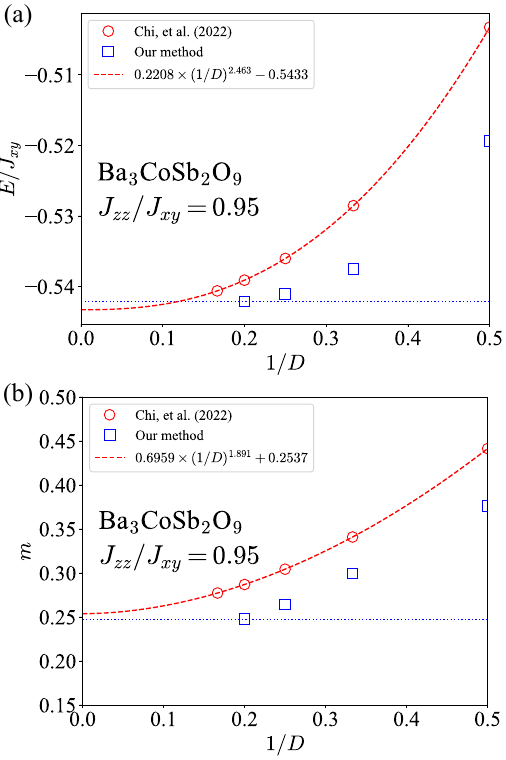}
    \caption{Comparison of (a) ground state energy and (b) magnetization for $\text{Ba}_3\text{CoSb}_2\text{O}_9$ with coarse-grained ansatz of Ref.~\cite{chi22-exci_xxz_trgl}.}
    \label{fig:scaling_comparison_bcso}
\end{figure}

In this section, we compare in detail our method using three-sublattice iPEPS ansatz to the coarse-grained ansatz used in Ref.~\cite{chi22-exci_xxz_trgl} from the perspective of both ground state properties and excitation spectra. We show the comparison of finite-$D$ data for the material $\text{Ba}_3\text{CoSb}_2\text{O}_9$ described by the easy-plane XXZ model on triangular lattice (Fig.~\ref{fig:scaling_comparison_bcso}).
Our ansatz shows faster convergence with respect to the bond dimension compared to the one used in Ref.~\cite{chi22-exci_xxz_trgl}. This can be understood in terms of entanglement -- for the coarse-grained ansatz used in Ref.~\cite{chi22-exci_xxz_trgl}, the auxiliary bond connects pairs of unit cells while in our ansatz the auxiliary bonds connect pairs of physical sites. Hence, our ansatz can saturate the area law of entanglement entropy~\cite{RevModPhys.82.277} at higher value with the same bond dimension and in effect it can sustain stronger and more distant correlations compared to the coarse-grained ansatz.

In particular, the order parameter $m$ is reduced from its mean-field value due to the influence of quantum correlations. These correlations are increasingly accounted for by tensor network as $D$ is increased and thus we only expect $m$ to monotonically decrease in optimized higher-$D$ networks. Our $D=5$ result, shown in Fig.~\ref{fig:scaling_comparison_bcso}(b), has the order parameter $m$ already below the extrapolated $D\to\infty$ value obtained in Ref.~\cite{chi22-exci_xxz_trgl} based on their coarse-grained ansatz. This result underlines the faster convergence of our ansatz with increasing bond dimension.

\begin{figure}[tbh]
    \centering
    \includegraphics[width=0.9\linewidth]{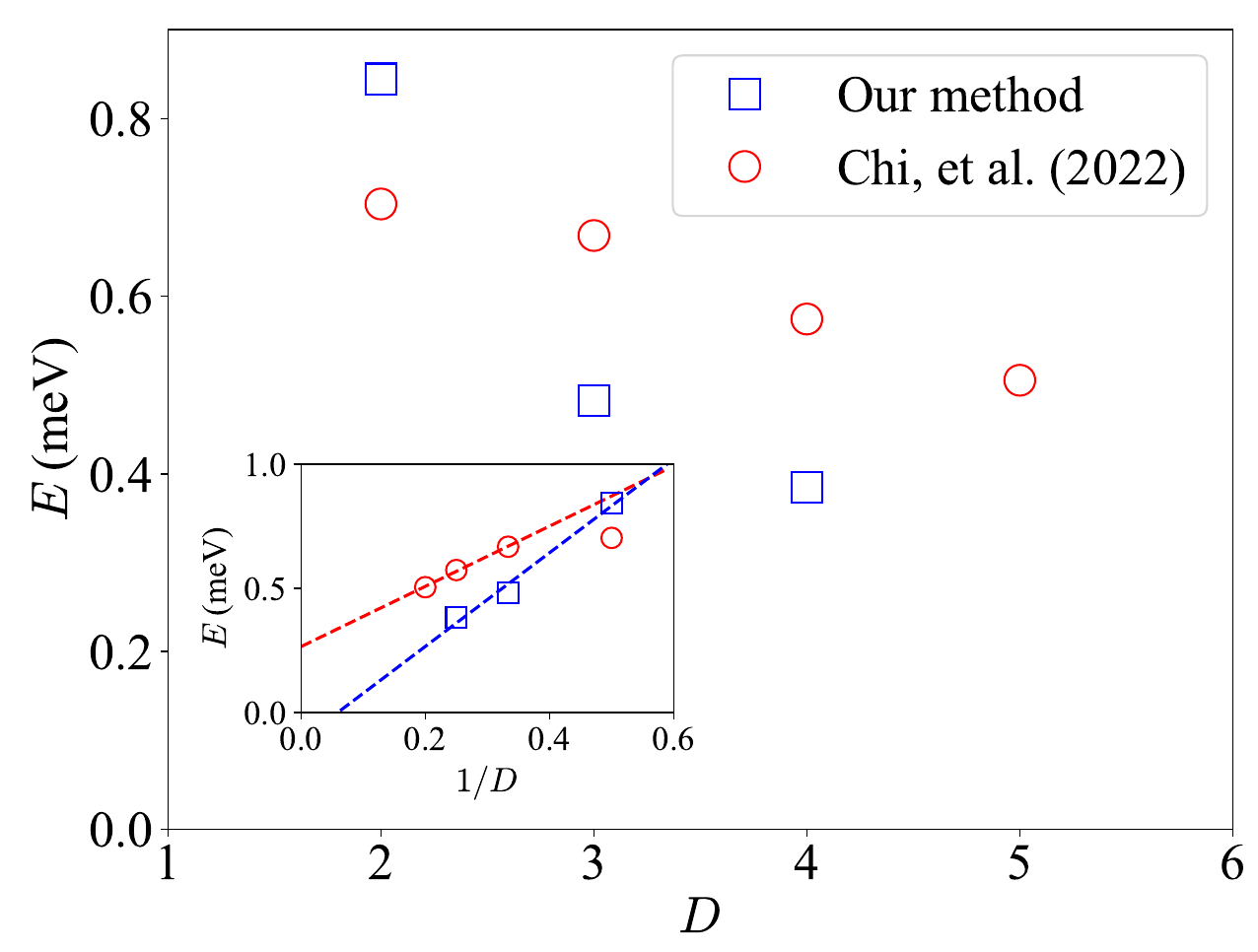}
    \caption{Comparison of the finite-$D$ scaling of the K-point gap for $\text{Ba}_3\text{CoSb}_2\text{O}_9$ with two different iPEPS ansatze -- the single-site anatz used in this work vs. the course-grained ansatz used in Ref.~\cite{chi22-exci_xxz_trgl}. Our ansatz shows faster convergence with increasing bond dimension, as it captures inter-site quantum fluctuations better.}
    \label{fig:kgap_scaling_bcso}
\end{figure}

In order to directly compare the excitation spectrum with Ref.~\cite{chi22-exci_xxz_trgl}, we compute the excitation spectrum at the $K$ point for $\text{Ba}_3\text{CoSb}_2\text{O}_9$. For this particular material described by the easy-plane XXZ model on triangular lattice, its zero-field ground state is described by a $120^{\circ}$ antiferromagnetic state where all three spins are aligned in the xy-plane with an angle of $120^{\circ}$ between the mutual spins. Similar to $\text{K}_2\text{Co}(\text{SeO}_3)_2$, such a configuration also spontaneously breaks the U(1) spin-rotational symmetry around the $S^z$-axis, giving rise to the gapless Goldstone mode around the  $K$ point. However, in practice, excitation PEPS ansatz with finite bond dimension $D$ cannot fully recover the gapless feature of the Goldstone mode because of its diverging correlation lengths. Hence, proper extrapolation to $D\to\infty$ limit is required. In Fig.~\ref{fig:kgap_scaling_bcso}, we show the $K$-point gaps for different bond dimensions. Again, we note that our ansatz has a much faster convergence with respect to the bond dimension $D$ compared to the coarse-grained ansatz. Also, in the inset of Fig.~\ref{fig:kgap_scaling_bcso}, we demonstrate the $1/D$ extrapolation.

\section{Excitation ansatz}

In this section, we derive how to properly compute the normalization and energies for the excited states, and most importantly, how to compute the dynamical structure factor, based on the excitation PEPS ansatz given by 
\begin{equation}
    \ket{\Phi_\vk(B)}=\sum_{\vx} e^{i\vk\cdot\vx}  \ket{\Phi_\vx (B)}
\end{equation}

\subsection{Excitation norm and energy matrices}
For two given excited states at a fixed momentum $\vk$, the overlap is given as follow.
\begin{align}
    \text{Norm}(B_\alpha, B_\beta)&\equiv\langle\Phi_\vk(B_\alpha)\ket{\Phi_\vk(B_\beta)}\notag\\&=\sum_{\vx,\vx'}e^{-i\vk\cdot\vx'}e^{i\vk\cdot\vx} \langle\Phi_{\vx'}(B_\alpha) \ket{\Phi_{\vx}(B_\beta)}\notag\\
    (\vx''=\vx-\vx')\rightarrow &=\sum_{\vx,\vx''}e^{i\vk\cdot\vx''}\langle\Phi_{\vx-\vx''}(B_\alpha) \ket{\Phi_{\vx}(B_\beta)}\notag\\
    &=\sum_{\vx,\vx''}e^{i\vk\cdot\vx''}\langle\Phi_{\vx=\vo}(B_\alpha) \ket{\Phi_{\vx''}(B_\beta)}\notag\\
    &=\sum_{\vx}\langle\Phi_{\vx=\vo}(B_\alpha) \ket{\Phi_{\vk}(B_\beta)}\notag\\&=N_s\langle\Phi_{\vx=\vo}(B_\alpha) \ket{\Phi_{\vk}(B_\beta)}
\end{align}
where $N_s$ is the number of sites. However, note that the definition above is actually an extensive quantity which diverges in the thermodynamic limits. Hence, in the actual implementation, the overlap of two excited states, $F^{e}_{\vk}(B_\alpha, B_\beta)$, is defined by the above overlap divided by the number of sites $N_s$, as given by Eq.~\ref{sup:eq:norm_mat}. With all the overlap between all possible pairs of excited states, one can form the effective norm matrix via $(\mathbb{F}^{e}_{\vk})_{\alpha\beta}\equiv F^{e}_{\vk}(B_\alpha, B_\beta)$. And one can easily verify that the matrix is Hermitian conjugate.
\begin{equation}
    F^{e}_{\vk}(B_\alpha, B_\beta)=\frac{1}{N_s}\langle\Phi_\vk(B_\alpha)\ket{\Phi_\vk(B_\beta)}=\langle\Phi_{\vx=\vo}(B_\alpha) \ket{\Phi_{\vk}(B_\beta)}
    \label{sup:eq:norm_mat}
\end{equation}

Similarly, one can compute the excited energy (relative to the ground state) matrix by sandwiching the Hamiltonian between two given excited states.
\begin{equation}
    (\mathbb{H}_\vk^e)_{\alpha\beta} \equiv E^{e}_{\vk}(B_\alpha, B_\beta)=\frac{1}{N_s}\langle \Phi_\vk(B_\alpha) | (\hat{H}-E_{gs}) \ket{\Phi_\vk(B_\beta)}
\end{equation}

Then, one can obtain the excited energy spectrum by solving the following generalized eigen-equation.
\begin{align}
    \sum_{\beta}(\mathbb{H}_\vk^e)_{\alpha\beta} u_{\beta}^{(\gamma)} = \sum_{\beta}E_\gamma(\mathbb{F}_\vk^e)_{\alpha\beta} u_{\beta}^{(\gamma)}
\end{align}

\subsection{Spectral weight}

In order to compute the dynamical structure factor as given by Eq.~6, the main goal is to obtain the spectral weight $p^\sigma_\alpha(\vq)$ for a given eigen-mode, which is obtained by solving the generalized eigen-equation above. Then, we simply insert the spin operator between the ground state and the given excited eigen-state at the fixed momentum $\vq$ (the momentum conservation is taken into consideration). This is equivalent to the procedure of computing the overlap between the excited state and the ground state with the spin operator acting on the ground state PEPS tensor, as shown in Eq.~\ref{sup:eqn:spec_weight}.
\begin{align}
p^\sigma_\alpha(\vq)&=\big|\langle\Phi_\vq(\tilde{B}_\alpha)|\hat{s}^{\sigma}_{\vq}\ket{\Psi(A)}\big|^2\notag\\
&=\big|\sum_{\vx'}e^{i\vq\cdot\vx'}\langle\Phi_{\vq}(\tilde{B}_\alpha)|\hat{s}^{\sigma}_{\vx'}\ket{\Psi(A)}\big|^2\notag\\&=\big|\langle\Phi_{\vq}(\tilde{B}_\alpha)\ket{\Phi_\vq(\hat{s}^\sigma\cdot A)}\big|^2\notag\\
&\sim \big|\langle\Phi_{\vq}(\tilde{B}_\alpha)\ket{\Phi_{\vx=\vo}(\hat{s}^\sigma\cdot A)}\big|^2\label{sup:eqn:spec_weight}
\end{align}
Then, the dynamical structure factor at the fixed momentum $\vq$ is simply a weighted sum of $\delta$-function peaks with the weights being the spectral weights computed above. In reality the density of states represented by these $\delta$-functions will be broadened due to finite-temperature effects. Furthermore, the neutron scattering spectra itself may have intrinsic continua caused by defects, inelastic scattering process, and quantum fluctuations such as magnon fractionalization in quantum spin liquid formation. In order to mimic the smearing behavior caused by both the finite-temperature effects and the finite experimental INS energy resolution, we apply a Lorentzian broadening to the spectral weights when computing the dynamical spin structure factor.

\end{document}